# ACCELERATION FOR THE µ STORAGE RING ν-SOURCE

David Neuffer, FNAL, Batavia, IL 60510, USA

*Abstract*

In a µ-storage ring neutrino source, a high-intensity proton source produces pions from a target, and muons from pi-decay are collected and cooled, accelerated to 10-50 GeV energies, and inserted into a storage ring, where mu-decay produces neutrino beams. The system must accept a large-emittance large-energy-spread beam. A low-frequency acceleration system is used to capture and phase rotate the µ's, and a ~200 MHz high-gradient system is used for cooling and initial acceleration. A superconducting rf (SRF) scenario using a 200 MHz linac, and 200 to 400 MHz recirculating linacs accelerates the beam to full energy. These systems, some variations and alternatives, and the current R&D status are discussed.

## 1 INTRODUCTION

In this paper we discuss the acceleration requirements of a µ-storage ring neutrino source, and describe the current accelerator research program. An overview of a µ-storage ring ν-source is shown in figure 1.[1] This example uses the parameters developed for the Fermilab feasibility study, which is a primary source for this report. The muon source is based on a high-intensity proton synchroton which generates short, high-intensity pulses of protons. These pulses are transported onto a high-intensity target, where they produce large numbers of pions. The pions are captured within a high-field solenoid, which is adiabatically matched to a low-field solenoid transport which also contains a low-frequency acceleration system. The pions decay to muons within the transport, while the acceleration system rotates the beam in phase space, limiting the energy spread while lengthening the bunches. The µ bunches are cooled by ionization cooling. The cooled bunches are then accelerated through a linac and a sequence of recirculating linacs to full energy, where they are inserted into a storage ring. µ-decay in the straight sections of the ring provide ν-beams, from (µ → e + $\bar{\nu}_e$ + $\nu_\mu$), which are directed toward a detector placed at a distance optimal for ν-oscillations (3000 km away in ref. 1). The design is largely based on the collaboration µ⁺-µ⁻ Collider studies [2, 3, 4] The ref. 1 design goal is a 50 GeV µ-storage ring, and uses a 1MW proton source (16 GeV, 15 Hz, $3 \times 10^{13}$ p/pulse) to obtain ~$3 \times 10^{20}$ stored µ/year or ~$10^{20}$ decays in a long-baseline straight section.

In this note we will describe the acceleration systems used for µ capture, cooling, and acceleration, with an emphasis on those used for acceleration of the beam to full energy. These acceleration systems are the largest cost items in the system and have some novel properties.

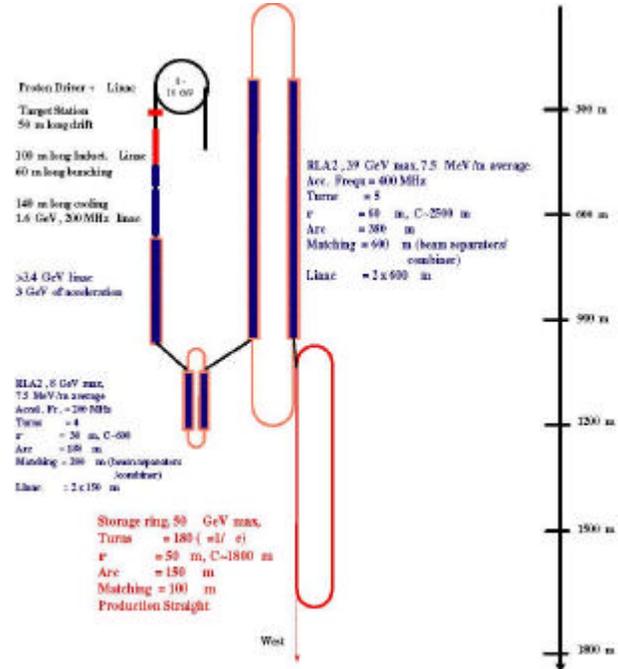

Fig. 1: Layout of a µ-storage ring ν-source. (from ref. 1).

## 2 ACCELERATION FOR µ CAPTURE

In the ν-factory scenario, protons produce π's in a target, and the resulting π-beam is captured in a solenoidal focusing system and continued through a transport where π's decay into µ's, following π → µ + $\bar{\nu}_\mu$. The initial π's, and the product µ's, are produced with a very large energy spread. Due to this energy spread, the beam bunch length increases in the decay transport and a longitudinal position-energy (φ-E) correlation develops with the lower energy beam trailing behind the higher energy muons. To reduce the energy spread, the beam passes through acceleration systems which decelerate the head of each bunch and accelerate the tail, obtaining a long bunch with small energy spread.

Two types of acceleration systems have been considered for this π→µ capture section:
- a low frequency rf system based on the use of ~30 MHz rf (~5 MV/m), and
- an induction linac system (~200 ns long pulses at ~1MV/m.

In the µ⁺-µ⁻ Collider scenarios of ref. [2,3] a low-frequency rf system is used, which requires a ~50m long ~30MHz rf system at ~5MV/m and obtains a ~10m long bunch with $\delta E_{rms}$~ 40MeV, which requires energy cooling for complete beam capture.

In the ν-source of ref. 1, a 50 m decay drift is followed by a 100m induction linac with ΔV = -0.5 to 1.5 MV/m is used, obtaining an ~60m long bunch with $\delta E_{rms} \cong 20$ MeV, $E_{kinetic}$ = 200MeV is obtained. This beam is transported through an absorber and a 200 MHz bunching system, which forms the ~10% energy spread bunch into a string of ~40 200 MHz bunchlets.

In ref [5] an rf system is combined with an induction linac, to obtain energy spreads minimized to 10 MeV or less, and from there captured into a string of 175 MHz bunchlets for subsequent cooling. Both the low frequency rf systems and the induction linac require significant extensions of present technology.

Designs for the induction linac module have been developed by S. Yu and I. Terechkine et al., and these initial designs are being developed. In ref. [1] the 100m long induction linac is composed of 1m modules (see fig. 2), with 0.1m gaps between conducting tubes. Each module consists of a set of toroidal magnetic cores surrounding an induction gap between cylindrical conductors. A pulser sends high-current pulses to the module that change the core magnetic fields, and the changing magnetic field generates an accelerating voltage across the induction gap. The induction module includes a 3 T solenoidal field, obtained from superconducting coils inside central conductors, inside the induction cores. The modules have a 20 cm radius beam channel within the conductors, and a 20 cm annular gap between the cylinder and the cores, so that the magnetic cores have an inner radius of 45cm and an outer radius of ~80 cm. The magnetic material of the cores can be Finemet, Metglas, or (newer) nanocrystalline compounds. Each module contains 10 6cm thick cores, so the entire system requires ~600000kg of material.

A significant problem is the development of a suitable power supply and pulser system for the induction linac. While single pulses are comparatively easy, Ref. 1 includes a design with 4 sequential pulses (~100 ns pulses spaced by 400 ns) at 15 Hz, corresponding to 4 initial proton bunches from a 15 Hz proton driver. For each gap, the inductor energy is supplied by a Thyristor switched capacitor bank (72 μF, 4kV), which feeds energy into 4 magnetic compressor pulse charge modules, 4 pulse forming lines, and delay lines and cell cables leading into the induction modules. The total system requires 8 MW cw power and is quite large in overall size and scope. Variations on this design concept are also being studied. Further design, as well as prototype testing, will be needed to find an optimal and verified solution.

Low-frequency rf is also possible and is more adapted to a multi-bunch scenario. However the cavities required are very large and the gradients needed are larger than previously obtained at these frequencies. Previous experience has been at 1—2 MV/m at these frequencies.

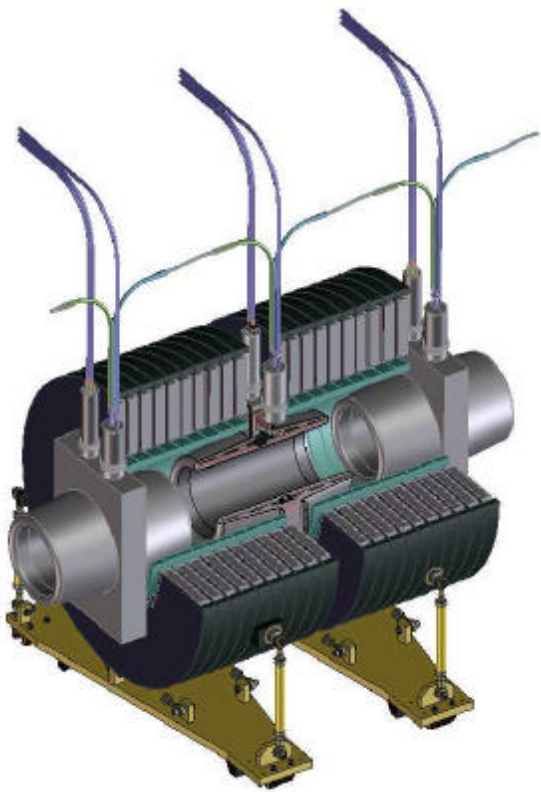

Fig. 2. Overview of two induction linac modules, showing cutaway views of cylindrical conducting tubes separated by accelerating gaps surrounded by ferrite cores, with focusing fields from solenoidal coils within the tubes.

## 3 ACCELERATION FOR mCOOLING

Following the φ-E rotation the μ-beam is captured into a string of rf buckets. To improve acceptance through the following accelerator and storage beam, The μ-beam is cooled using ionization cooling. The baseline cooling requirement is a reduction of transverse emittances by a factor of ~10, to $\varepsilon_{t,N,rms} \cong 0.0015$ m.

The cooling system requires high gradient acceleration as well as strong focusing. In the development of the ν-source studies, a ~ 200 MHz frequency was chosen. The frequency is high enough that high power sources are practical, and the cavities are large enough to accommodate the large emittance beams. To confine the large emittance beams and to focus the beams into the absorbers, strong solenoidal magnetic fields with B ≅ 5T are needed, which include coils surrounding the cavities. Since superconducting rf (SRF) does not function in high magnetic fields, it is not used here.

Since μ's interact weakly with materials, it is possible to have cavities with material crossing the beam line. One cavity design under construction consists of cylindrical "pill-box" Cu cavities with Be foils covering the beam irises. In this configuration the maximum electric fields are at the acceleration gaps, which enables maximum acceleration gradients. An alternative confi-guration has a gridded iris. The cooling system requires acceleration gradients of ~15 MV/m at 200 MHz.

Figure 3 shows a cooling cell with 200 MHz gridded cavities, liquid hydrogen absorbers, and superconducting magnetic coils providing ~5T focusing fields. ~100m of such structures are required in a complete cooling system.

RF power requirements are quite high for this normal-conducting, high-gradient, low-frequency structure. A multibeam 10 MW klystron [1] is required for every ~2m of cavity structure, or ~75 klystrons for a complete 150m long cooling channel. Initial designs are ~4m long multibeam klystrons, with ~2.4m long IGBT modulators; these coud be arranged transverse to the cavities. The overall power requirements are quite large: ~750MW peak power but "only" ~2MW average power.

While various other possible cooling configurations are under consideration, the major variants are in the external focusing systems and in the absorber configurations. The rf requirements for the required cooling remain roughly the same: ~100m (or more) of ~15MV/m, ~200MHz rf acceleration structures.

However, the total installed cooling rf could be reduced if a multiturn cooling system were developed. A "Ring Cooler" with up to 10 turns may be possible.

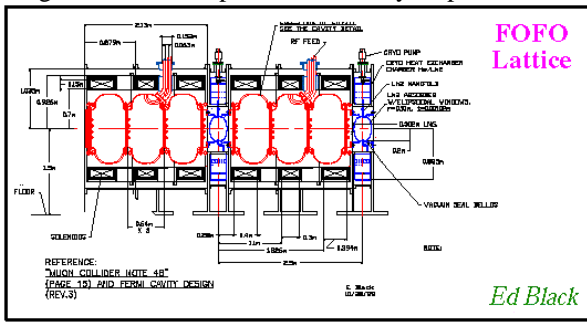

Figure 3: A cell for µ-cooling, including gridded acceleration cavities, liquid hydrogen absorbers, and magnetic coils.

## 4 ACCELERATION SCENARIOS

A baseline acceleration scenario was developed for ref. 1,[6] and after some modification and development was included in that study. Parameters are displayed in Table 1. It consists of a 3 GeV Linac, which injects into a 4-turn RLA (using 200 MHz SRF) which would take the beam to 11GeV, followed by a 5-turn RLA to take the beam to 50 GeV (using 400 MHz SRF). The design concept has been explored in some detail by Lebedev, who has developed beam transport systems for the linac, and by J. Delayen and H. Padamsee, who explored acceleration requirements.[1]

For final µ energies greater than ~10 GeV the acceleration system dominates the cost of the ν-Factory, and the acceleration structures and power sources dominate that cost. Multiturn use of the acceleration structures (in an RLA) reduces the rf costs (balanced by an increase in beam transport costs). Also higher-frequency SRF should be much lower in cost (both in structure and power requirements), and this mandates an increase in frequency where the beam dynamics allows it.

The critical requirement is that the acceleration must be fast enough to avoid µ-decay, which means real-estate acceleration gradients more than a few MeV/m; say eV′ > ~5MeV/m. This includes transport elements; the design goal is V′ > 15 MV/m within the acceleration cavity and < 67% of the transport length devoted to non-cavities (i. e., focusing, return arcs, spreaders, cryo-module connections, etc.).

### 4.1 Linac

Initial acceleration is obtained in a 200 MHz SRF linac. In the linac, the initial acceleration is off crest (by −77°)in order to capture the beam from cooling, which has an rms bunch length of ~15cm (5σ full width = 0.75m or 180°) and rms energy width of 20MeV (100MeV full width). As the bunch accelerates the beam adiabatically damps over the first ~1 GeV of acceleration to $\sigma_z \cong$ 4cm and $\delta E_{rms} \cong$ 80 MeV. The phase is moved to crest over the linac length (a linear ramp in phase with distance was used).

In Ref. [1, 9] transverse focussing from injection to ~ 2 GeV is provided by solenoids, with central field increasing from ~1T to 5T over the linac. Initially the solenoids are spaced between two cavities (~6m apart), and this is increased to 4-cavity cells at 1 GeV (~12m cells) Throughout the linac the accelerating cells occupy ~50% of the actual linac length, so the effective accelerating gradient is ~7.5 MV/m. There may be diffi-culty in isolating the solenoidal fields from the superconducting cavities.

Solenoid focusing is not essential in the linac. Popovic has developed a scenario relying on quadrupole FODO focusing, which could be continued into the RLA.[8]

For the remainder of the linac, and for the following RLA linacs and arcs, a triplet focusing configuration is used (in ref. 1).[9] The advantages of a triplet is that the beam remains nearly round throughout the cavities (focused in both x and y), enabling long magnet-free sections for multi-cavity cryomodules (up to 10m long). Also only one dimension (i. e., x) has a large size in the triplet, and therefore large chromaticity; a 1-D chromatic correction may be adequate. Alternative "FODO" lattices for the arcs, which enable both x and y correction are also possible.

### 4.2 RLA design

Linac acceleration is continued until the beam is fully relativistic and the longitudinal beam size (δz-δE/E) is matched into a RLA acceptance. For ref. 1, 3 GeV was chosen for the transition energy into a first RLA.

The principles of µ-RLA acceleration are described in ref. [7]. In an RLA the beam is injected into a linac, accelerated, and returned by arc transports for multiple passes of acceleration through the same linac(s). At the end of each linac, beam transport (through dipoles) sorts the beam by energy directing each pass beam through a separate return arc transport. At the end of the arc the separate transports are recombined for acceleration in the following linac. To cleanly separate the various passes, the beam energy width must be less than the energy difference. The energy widths and the costs of separate transports limit the number of recirculation turns.

Because µ-bunches have a large longitudinal phase space, it is desirable to accelerate off-crest and use non-isochronous return arcs to obtain stable longitudinal motion (controlling energy spread δE) through each pass as well as to match the $\sigma_z$-$\delta E$ bunch sizes into the following accelerators. Since each return transport is separate the central phase $\phi_s$ and chronicity $M_{56} = d(\delta z)/d(\delta p/p)$ can be tuned in each turn.

The first RLA (200 MHz) takes the beam from the linac to a suitable energy for injection into RLA2 (400MHz). The longitudinal motion must confine the beam and match the beam into RLA2, which means reducing the bunch length by a factor of ~2. The beam transport must accept the full energy spread, which changes from $\delta E/E = \pm 2.5\sigma = \pm 8\%$ to $\pm 4\%$ at extraction into RLA2, with the rms bunch length changing from 4.2 to 2.2cm, as the beam is accelerated from 3 to 11 GeV. Lattices with the desired optics have been developed by Lebedev and Bogascz.[9]

In RLA2 the beam is accelerated within a stable rf bucket and then injected into the storage ring, where the design constraint is to obtain a minimal energy spread for easier acceptance in the storage ring. At the parameters of table 1, the energy spread reduces from ±4% to ±1% (with a maximum of ±6% at E = 20GeV) in RLA2 with a maximum of as the beam accelerates to 50 GeV. A similar scenario of Lebedev[9] uses $M_{56}$ = 2 m and keeps the beam energy small throughout RLA2, at the cost of a smaller acceleration bucket. Transverse motion is not yet detailed for RLA2. Results of simulations of the longitudinal motion through the acceleration sequence are displayed in fig. [5].

*4.3 SRF and power considerations.*

The 200 MHz cavities are 1.5m long two-cell cavities. Scaling from TESLA/LEP II, the cells are 75cm long, and have an outer radius of 65 cm with beam apertures of ~20cm. The 400 MHz cavities are half the transverse size of these, and would come in 4-cell 1.5m cavities (similar to LEP II). These would be grouped into cryo-modules consistent with the focusing; 10m (4-cavity) cryomodules are in the present design.

RF power design is also discussed in ref. [1]. The 200 MHz cavities in the linac and RLA1 require 820 kW/cell with a fill time of 2ms. The 400 MHz cavities of RLA2 require only 200 kW/cell. With klystrons of peak power of 8—9MW (240—270kW ave. power), the linac requires 32 klystrons (10 cells/klystron, 3.6GV total voltage), RLA 1 requires 23 klystrons (2.6 GV total voltage), and RLA2 requires 25 klystrons (60 cells/klystron, 8.5 GV total) The total average power is 20.4 MW (7.9, 5.7, 6.8 MW for Linac, RLA1, RLA2). Note that a 200 MHz RLA2 would require 4× more power. The total number of klystrons in the SRF accelerator is approximately the same as that of the normal conducting cooling section, which has less than 2GV of total voltage.

*4.4 Acceleration variations*

Since the accelerator is the largest cost item, it can be possible to greatly decrease the cost by changes in that system. It would be desirable to use higher-frequency rf systems, and a scenario which moved the acceleration from 200 to 400 MHz earlier (and possibly on to 800 …) could be much more affordable. More cooling in the scenario (including energy cooling) may enable this.

Overall costs could also be reduced by adding more passes in each RLA, reducing rf requirements at the cost of additional transport.

Significant changes are also required if a different final energy is chosen. For a 20 GeV scenario, a 2 GeV linac with a single 6-pass RLA is being considered. A 400 MHz rf system could be used for the RLA, if beam size and energy spread are controlled.

Costs could also be reduced if an "FFAG-like" configuration could be developed. This would be similar to an RLA with acceleration modules and return arc transport. The difference is that the beam transport is designed with a broad enough energy spread acceptance that all of the energy turns return within the same arc. Thus the number of turns of acceleration could be increased from 4—5 to ~10—20, reducing the rf structure requirements. The transport costs may be significantly less (since the number of return arcs is reduced, but at the cost of higher δE-acceptance transports).

## 5 RF R&D PROGRAM

The µ-storage ring ν-source requires new rf systems and R&D programs have been initiated to develop these systems. The minimum rf system requirements are an induction linac (or low-frequency rf) for initial capture, a ~200 MHz normal-conducting rf system for cooling, and an efficient 200 MHz SRF acceleration rf. R&D on development of induction linac systems has been initiated by S. Lee at LBL. 200 MHz rf studies are based at LBL and at Fermilab. Initial designs of rf cavities have been developed, both with Be windows and with gridded irises. These studies will build on the 805 MHz R&D program, which has built Be window cavities (and open cell cavities), and will provide high-gradient tests of these structures. A test setup for 200 MHz rf will be added to the Fermilab 805 MHz test lab. Also Super-conducting rf programs designing and building prototype accelerating cavities have been initiated at Cornell and at TJNAF.

## 6 CURRENT STATUS AND PLANS

A 6-months study on the physics potential[10] for a µ-storage ring ν-source and a feasibility study on the beam facilities[1] have just been completed at Fermilab. These studies conclude that the physics potential of that source is interesting and important, and that the general concept is technically feasible. The feasibility study identified key "cost drivers" (expensive components and features) and performance limitations; the critical cost drivers are the acceleration and beam cooling systems.

In the next year, a design study based at BNL will be developed with somewhat different scope, and facility requirements, and some site-dependent constraints. In this study the collection and cooling channel will be reoptimized for performance and affordability, and the acceleration will also be reoptimized.

**50 GeV µAccelerator Parameters**

| Parameter | Linac | RLA1 | RLA2 |
|---|---|---|---|
| Initial Kinetic Energy | 0.12 | 3 | 11 GeV |
| Final Kinetic Energy | 3.0 | 11 | 50 GeV |
| Number of turns | 1 | 4 | 5 |
| Rf Voltage/linac | 3.6 | 1.15 | 4.3 GV |
| Acceleration phase | 77→-6° | 30° | 25→0° |
| M$_{56}$ per arc | --- | 0.5→1.4 | 0.3→0 m |
| Rf frequency | 200 | 200 | 400 MHz |
| Rms bunch length | 15 → | 4 → 2 | → 1.5 cm |
| Rms energy spread | 20 → | 80 → 160 | → 220 MeV |
| Linac length (2 per RLA) | 500 | 200 | 800 m |
| Arc length | --- | 220 | 700 m |

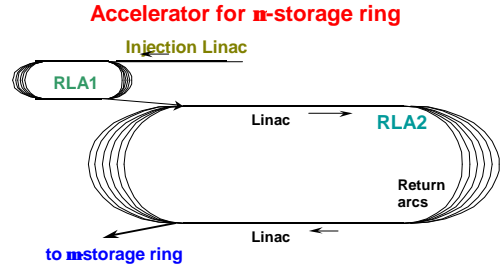

Fig. 4. Overview of accelerator for the µ-storage ring ν-source.

Fig. 5 Simulation of longitudinal motion in acceleration; φ-δE plots of beam at linac injection (0.12 GeV kinetic energy), beginning of RLA1 (2.97GeV), beginning of RLA2 (10.9 GeV, 400 MHz), and end of RLA2(50.7 GeV).

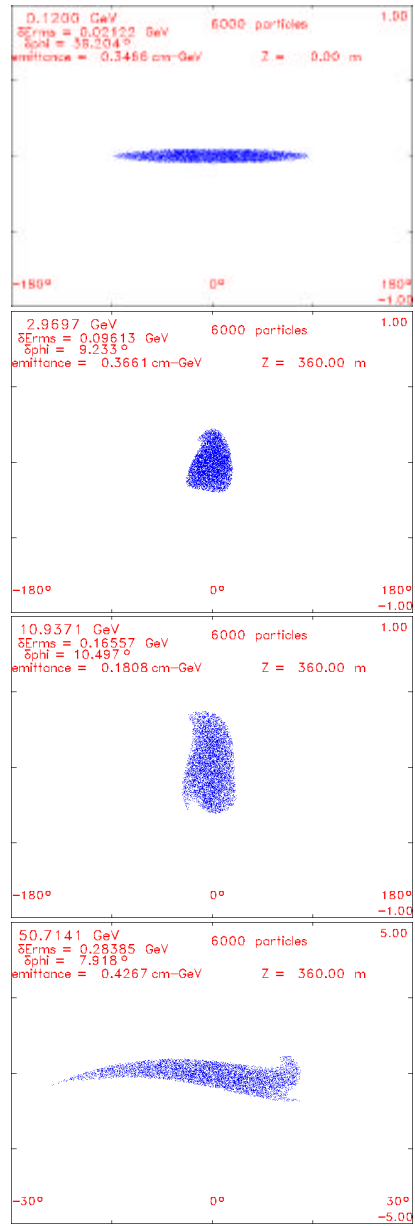